# Controlled synthesis of THz beams spanning the higher-order Poincaré sphere

**YAQUN LIU*, VALDAS PASISKEVICIUS**

*Department of Applied Physics, KTH Royal Institute of Technology, Roslagstullsbacken 21, 106 91 Stockholm, Sweden*
*yaqun@kth.se

## Abstract

Terahertz (THz) structured fields comprising separable and nonseparable spin and orbital angular momentum states offer unique opportunities for light-matter interactions in chiral media and material systems containing topological electronic spin excitations. This requires deterministic synthesis and control of topological vector vortex states that can be conveniently mapped onto the two-dimensional higher-order Poincaré (HOP) sphere. Here, we establish a nonlinear method for synthesizing vector vortex THz beams corresponding to the HOP states using optical rectification driven by a pair of infrared pump pulses with polarization singularities. The method provides two independent control parameters that enable flexible access to arbitrary states on the HOP sphere. The resulting mapping is experimentally validated using representative states, covering linear, circular, and elliptical regimes, in excellent agreement with theoretical predictions. Our approach establishes a two-parameter control framework for structured THz field synthesis, providing a physically transparent and scalable route to topological engineering in the THz regime.

## Introduction

Structured light fields with phase or polarization singularities, such as vortex beams, Bessel beams, and light springs, have attracted significant attention [1-7]. Among these, cylindrical vector (CV) beams constitute a class of robust structured field states with coupled spin angular momentum (SAM) and orbital angular momentum (OAM) [8,9]. More generally, structured beams that simultaneously carry SAM and OAM are referred to as vector vortex beams [10]. Such beams provide additional degrees of freedom for encoding information and manipulating light-matter interactions [11,12]. Vector vortex beams can be mapped onto the higher-order Poincaré (HOP) sphere [13,14], forming an effectively two-dimensional parameter space in which arbitrary states can be represented as superpositions of CV beams with controlled relative amplitudes and phases. This framework is closely related to geometric phase effects, including the Pancharatnam-Berry phase arising from cyclic evolution on the sphere [15,16]. Various approaches have been developed to generate such states, including the use of Hermite-Gaussian modes [8] and polarization-conversion elements such as q-plate or S-plates [10,17-19].

In the terahertz (THz) regime, such structured fields are particularly attractive for applications in spectroscopy [20], wireless communications [21], and the control of chiral and magnetic excitations in matter [22-24]. Despite these opportunities, achieving deterministic and flexible control of THz vector vortex beams remains challenging. Conventional approaches based on external conversion elements are often bulky due to the large wavelength scale in the THz regime [25,26]. Alternative approaches based on metasurfaces rely on complex fabrication [27,28]. These elements all introduce additional insertion loss and typically enable only a limited subset of states on the HOP sphere. In contrast, direct generation of THz fields via nonlinear optical processes avoids the need for external THz structuring elements, providing an efficient route for structured field synthesis. For example, Lin *et al.* demonstrated the ability to tailor the wavefront of THz radiation using a dual-pump difference-frequency generation scheme, enabling the generation of THz scalar vortex beams carrying only OAM [29]. Optical rectification (OR), as another widely used nonlinear optical process in THz generation, has been employed to generate vortex THz beams using structured infrared pumps in high-symmetry nonlinear crystals [30,31]. The dynamic and transverse spatial phase of the pump field is not directly transferred to the THz field due to the self-difference-frequency nature of OR, making the direct transfer of phase singularities nontrivial. A general approach for synthesizing arbitrary vector vortex THz beams with simultaneous control of both OAM and SAM has therefore been lacking.

Nevertheless, the intrinsic phase stability of OR and its polarization-insensitive nonlinear response along the three-fold symmetric axis of the zinc-blende nonlinear crystal suggest a viable route toward simultaneous control of both the global phase structure and local polarization of the generated THz field. Here, we demonstrate a nonlinear synthesis platform for generating arbitrary vector vortex THz beams on the HOP sphere using two structured pump pulses containing singularities interacting along the three-fold rotation axis of the nonlinear crystal based on OR. By introducing two independent control parameters, the temporal phase delay between dual pump pulses and the rotation angle of an S-waveplate, we establish a direct and predictive mapping between the pump configuration and the resulting THz polarization states, spanning the full HOP sphere. This framework enables deterministic access to arbitrary vector vortex states, providing a powerful route for structured THz field engineering.

## Results

**Two-parameter mapping of THz vector vortex states.** To establish deterministic control of the generated THz vector vortex states, we consider two independent parameters defined by the pump fields. The phase delay φ between the pump pulses and the rotation angle θ of the S-waveplate define the two control parameters.

A theoretical framework provides a deterministic mapping between the control parameters and the resulting states (see **Supplementary Information**). The synthesized vector vortex THz field in the reconstruction coordinate frame (see **Methods**) can be described by equation (1).

$$P = \left[\cos\left(\frac{\varphi}{2}+\frac{\pi}{4}\right)e^{\left(+i2\phi-\frac{\pi}{2}-2\theta\right)}\right]\left|L\right\rangle + \left[\sin\left(\frac{\varphi}{2}+\frac{\pi}{4}\right)e^{\left(-i2\phi-\frac{\pi}{2}-2\theta\right)}\right]\left|R\right\rangle \quad (1)$$

Here, |L⟩ and |R⟩ denote the left- and right-handed circular polarization basis states, respectively, and the coefficients determine the relative amplitude and phase that define the resulting vector vortex state. ϕ represents the spatial azimuthal phase of the beam profile, associated with OAM. As established by the theoretical model, (θ, φ) can be conveniently expressed as a pair of spherical coordinates on the HOP sphere, where φ defines the polar angle and θ specifies the azimuthal angle. Within this framework, the resulting THz polarization states can be represented on the HOP sphere determined by the parameter pair (θ, φ). As illustrated in Fig. 1, systematic variation of these two parameters gives rise to a structured parametric grid on the HOP sphere, enabling full coverage of the state space of vector vortex beams. The beams on equator have spiral linear polarization structure which is known as CV beams [8]. Here, φ governs the ellipticity of the local polarization, while θ determines its azimuthal orientation distribution on the HOP sphere. This two-parameter representation provides a physically transparent description of the nonlinear generation process and establishes a direct link between experimentally controllable pump conditions and the synthesized THz field structure. All HOP states represent vector field with spatial singularity. The singularity can be characterized by a topological winding number l, the value of which represents how many times of 2π the azimuthal phase/polarization acquires counter-clockwise around the singularity. As can be seen in Fig. 1, all HOP states have constant absolute value of winding number |l| equal to 2, which is set by the OR doubling rule for the S-waveplate used here [30] while the signs of the SAM and OAM are controlled by the pump configuration.

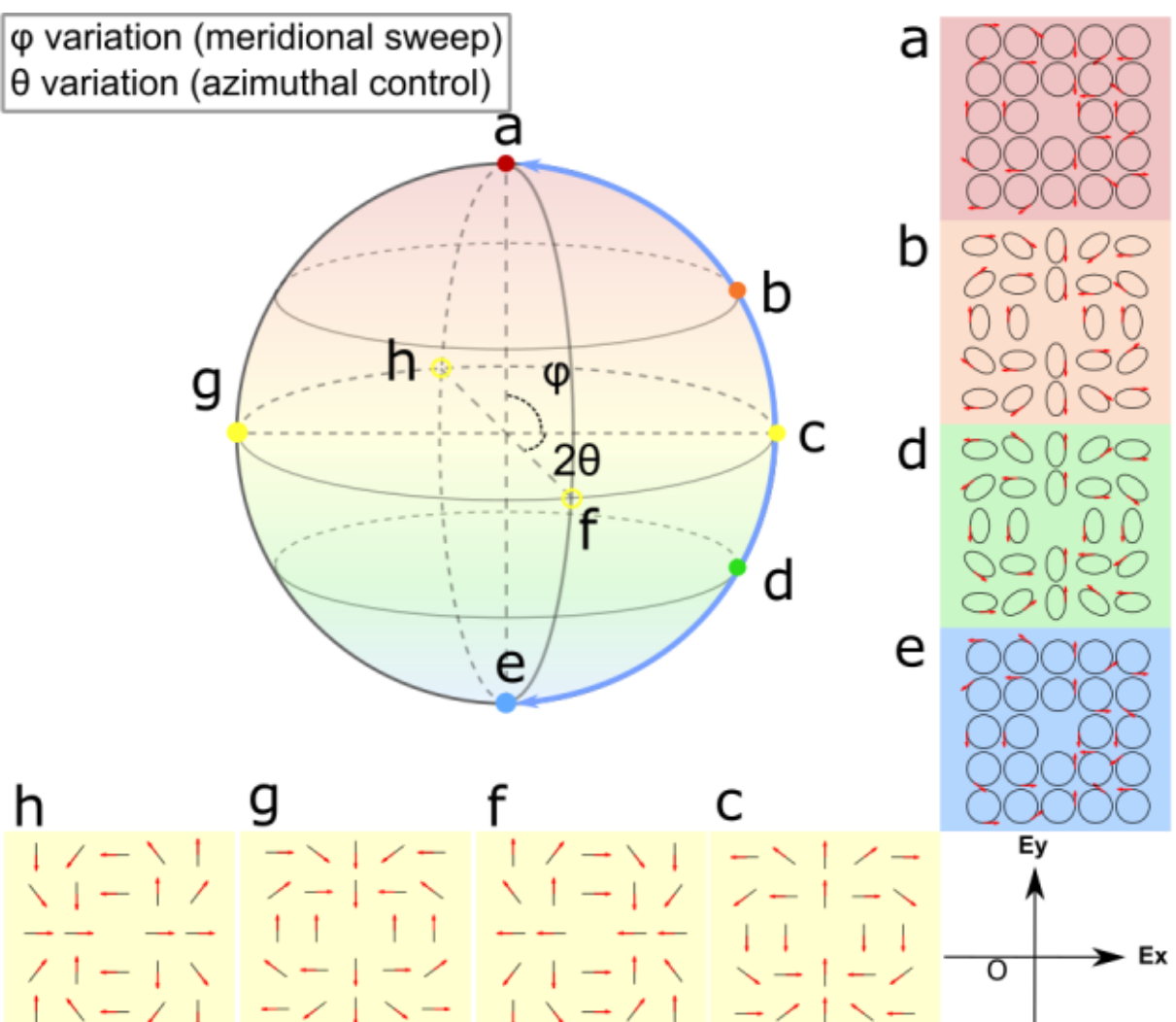

Fig. 1. Scheme of higher-order Poincaré sphere corresponding to |l|=2. The two independent parameters φ and θ define the polar and azimuthal angles, forming a structured mapping over the sphere. Variation of φ traces a meridional trajectory (blue curve), while θ determines the azimuthal position along the equator (yellow markers). Solid markers denote experimentally verified states, while open markers indicate additional accessible states not experimentally measured. The insets represent state of polarization (SOP) of the marked states. **a**: right-hand circular polarized (red), **b**: right-hand elliptical polarized (orange), **d**: left-hand elliptical polarized (green), **e**: left-hand circular polarized (light blue), **c, f, g** and **h**: linearly polarized CV states at azimuthal positions θ = 0°, 45°, 90°, and 135° on the equator (yellow).

**Experiment implementation.** The two control parameters introduced are experimentally implemented through the relative temporal control between the pump pulses and the polarization configuration defined by the S-waveplate. To illustrate the underlying mechanism, we first consider the local THz field at a fixed spatial position. As illustrated in Fig. 2**a,** the polarization state of the generated THz field can be continuously controlled by introducing a relative phase delay between two linearly polarized pump pulses with 45° polarization offset. A phase delay of π/2 in targeting THz frequency yields right circular polarization, arising from the superposition of two orthogonal THz field components with a quarter-cycle phase offset. The phase-dependent circularly polarized THz waveform is illustrated in Fig. 2**a** by simulated temporal signals, confirming the controllable generation of distinct polarization states. Similarly, a phase delay of -π/2 results in left circular polarization of the THz signal.

This mechanism is implemented experimentally using the setup shown in Fig. 2**b**. The pump beam is split into two arms with independently controlled polarization and relative delay, implemented using a delay line. The zero-phase delay between the two pump pulses is calibrated using two delay lines (see **Methods**), ensuring accurate control of the relative phase φ. The two pump beams are subsequently converted into CV beams using an S-waveplate prior to the nonlinear crystal. They are spatially overlapped in a ZnTe crystal for THz generation process (see **Methods**). The 2D distribution of the THz amplitude and transversal phase is characterized using polarization-resolved electro-optic sampling (EOS). The orthogonal THz field components $E_x$ and $E_y$ are measured to reconstruct the state of THz field based on the THz polarimetry technique (see **Methods**). The CV beam profiles and polarization states of the pump beams at the initial state (state **c** in Fig. 1, with both control parameters set to zero) are first experimentally verified to establish well-defined starting conditions for the subsequent mapping of vector vortex states.

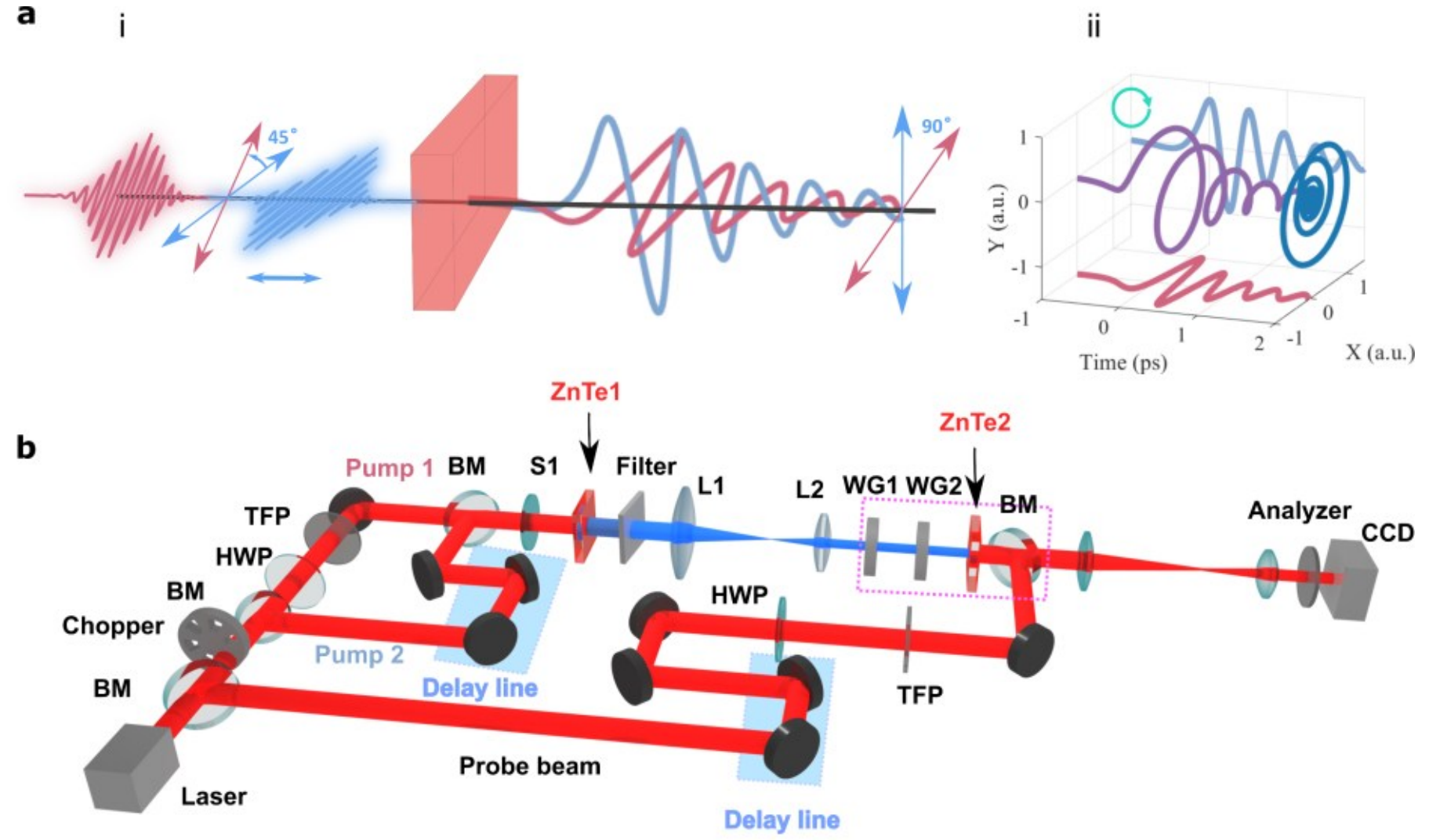

Fig. 2. Experimental implementation of THz vector vortex beam generation. **a**. Schematic illustration of the local THz field generation mechanism. i, Dual pump pulses with a phase delay of φ = π/2 (red at a phase of 0 and blue at a phase of π/2) results in circular polarization. Arrows: represent polarization orientation; ii, THz trajectory with right circular polarization; purple: waveform trajectory, Navy: waveform projection, red and blue: two perpendicular THz electric field components, cyan circle arrows: handedness. **b**. Experimental setup for THz generation and characterization. BM: beamsplitter mirror, HWP: half waveplate, TFP: thin film polarizer, S1: S-waveplate, ZnTe1: 111-ZnTe, ZnTe2: 110-ZnTe. WG1,WG2: wire grid polarizer.

**Experimental realization of THz states across the HOP sphere**. First, five typical states of complex vector THz vortices illustrated on a meridian indicated in blue in Fig. 1 **a-e** are generated by manipulating the phase delay between two pumps with S-waveplate setting at 0°. The experimental results are measured and characterized to prove the validity and feasibility of phase delay parameter φ of this scheme.

A phase delay φ of **π/2** was set by adjusting the delay-line in Pump 2 arm as shown in Fig. 2(b). To implement the reconstruction of THz field, the decomposition into two orthogonal eigenstates in the linear polarization basis - $E_x$ and $E_y$ was introduced, of which the orientations are shown in **Methods**. To visualize the evolution of THz field, the temporal electrical field distributions of $E_x$ and $E_y$ were imaged in a range of 1.6 mm, with a step of 0.01 mm, that in the time range of 5.3 ps with a resolution of 33 fs, separately. Fourier transform was subsequently employed to map the amplitude and phase distributions of these two eigenstates. The amplitude and phase distributions of $E_y$ at 0.95 THz are shown in Fig. 3(a) and (c), while that of $E_x$ are shown in Fig. 3(b) and (d). The profiles in this work are shown from the point of viewer. The spiral phases depicted in Fig. 3(c) and (d) both exhibit an anticlockwise rotation of 4π indicating the presence of OAM of -2ħ. These two linearly polarized eigenstates are revealed to be two scalar vortex beams with the same topological charge (TC) of -2. The simulated reconstructed state of polarization (SOP) of the THz field is shown in Fig. 3(e). A typical experimental reconstructed 3D THz local waveform trajectory and projection of the beam position at [0.9 mm, 0.9 mm] are shown in Fig. 3(f), corresponding to the simulated trajectory marked in red in Fig. 3(e). The handedness in the figure confirms the right-handedness, which is marked in Cyan circle in Fig. 3(f), exhibiting the SAM of -ħ. Note that the handedness of the experimental results demonstrated in this work are from the point of view of the observer. This beam corresponds to the SOP of Fig.1-**a** on the North pole illustrated in light red.

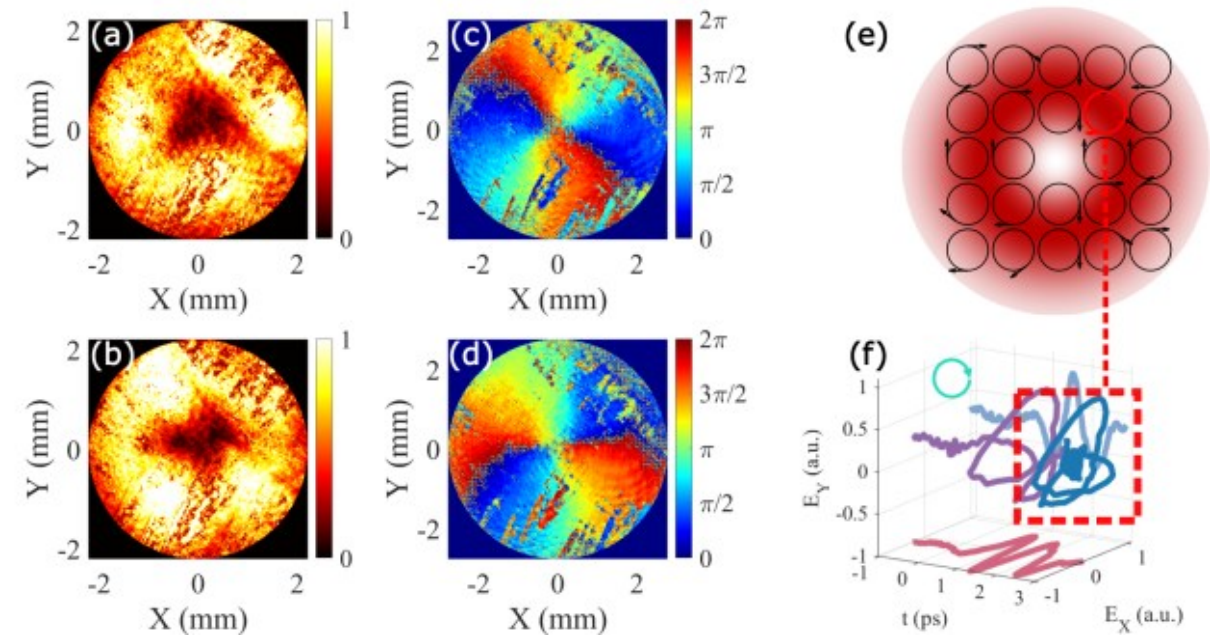

**Fig. 3. Measurement of the THz vortex for l = -2, s=-1. (a)**, THz electric field amplitude distribution along $E_y$ axis. (**b**), Amplitude distribution along $E_x$ axis. (**c**), Phase distribution along $E_y$ axis. (**d**),

Phase distribution along $E_x$ axis. Color bars in (**a**) and (**b**) show the intensities in arbitrary units, and color bars in (**c**) and (**d**) represent the phase scale (radian). (**e**), Simulated trajectory of THz temporal waveform. (**f**), measured trajectory of local THz temporal waveform (the position labelled red in SOP (**e**)).

By controlling the pulse delay between the pump beams, the polarization handedness of the vortex beam can be flipped from the right-circular to left-circular polarization. Fig. 4 shows the experimental results of two eigenstates of the left circular polarized THz vortex, corresponding to the South pole HOP state of Fig. 1-**e**. Similar to the first case, the measured 3D electric field trajectory of the same position is shown in Fig. 4(f), confirming the left-handedness. This handedness exhibits the SAM of +ħ. The phase delay reverses the handedness of the circular polarization and, simultaneously, flips the sign of TC. The phases in Fig. 4(c) and (d) both show a 4π clockwise spiral feature, indicating an OAM of +2ħ.

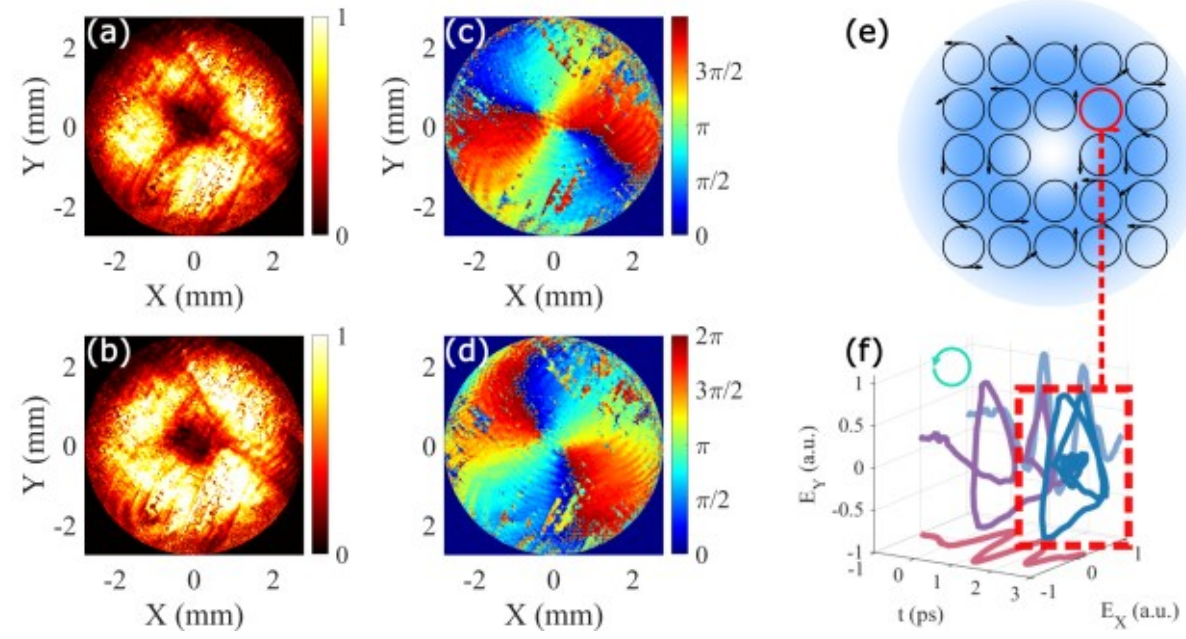

**Fig. 4. Measurement of the THz vortex for l = +2, s=+1. (a)**, THz electric field amplitude distribution along $E_y$ axis. (**b**), Amplitude distribution along $E_x$ axis. (**c**), Phase distribution along $E_y$ axis. (**d**), Phase distribution along $E_x$ axis. Color bars in (**a**) and (**b**) show the intensities in arbitrary units, and color bars in (**c**) and (**d**) represent the phase scale (radian). (**e**), Simulated trajectory of THz temporal waveform. (**f**), measured trajectory of local THz temporal waveform (the position labelled red in SOP (**e**)).

The local polarization can be controlled between circular, elliptical, and linear polarization depending on the pulse phase delay between two pump beams. Among all the beams, the general case is the vortex beam with local elliptically polarizations. The vortex beams with right and left elliptically polarization, of which the SOPs are plotted in orange and green on the HOP in Fig. 1, were also generated and characterized.

The experimental results of THz vortex beam were measured while the phase delay φ was set as **π/4** as shown in Fig. 5. The evolutions of wavefront of the two eigenstates over 1.65 ps (50 frames) are shown in the attached videos (Fig. 5_Ex.mp4 and Fig. 5_Ey.mp4). The phase distributions in Fig. 5(c) and (d) verify a TC of -2. However, the amplitude profiles in Fig. 5(a) and (b) are not as uniform as that in Fig. 3 and Fig. 4. This non-uniformity results from the local elliptical polarization and represents the expected conoscopic picture revealed through polarimetric measurement. That picture will be even clearer for linearly polarized HOP equator states. The amplitude of $E_y$ in Fig. 5(a) has maximum values along horizontal and vertical directions through the beam center, while the $E_x$ amplitude in Fig. 5(b) has minimum value along horizontal and vertical directions. The non-uniform feature is consistent with the SOP shown in Fig. 5(e) as we expected. For the local elliptical polarizations that along horizontal and vertical directions through the beam center, the long axis is along $E_y$ while the short axis of that is along $E_x$. The measured THz trajectory in Fig. 5(f) corresponds with the simulated in red in Fig. 5(e), that are right-handed and elliptically polarized with long axis along $E_x$. As a general beam on the HOP sphere of Fig.1-**b**, it further verifies experimental agreement to the theory.

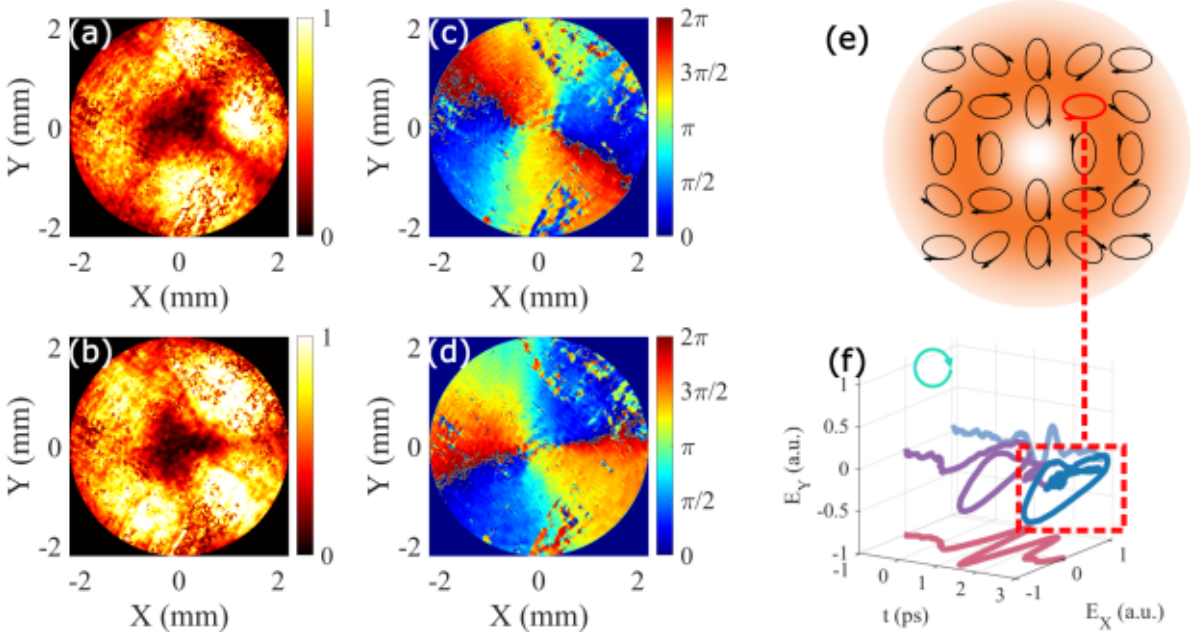

**Fig. 5. Measurement of right-handed elliptical THz vortex beam. (a)**, THz electric field amplitude distribution along $E_y$ axis. (**b**), Amplitude distribution along $E_x$ axis. (**c**), Phase distribution along $E_y$ axis. (**d**), Phase distribution along $E_x$ axis. Color bars in (**a**) and (**b**) show the intensities in arbitrary units, and color bars in (**c**) and (**d**) represent the phase scale (radian). (**e**), Simulated trajectory of THz temporal waveform. (**f**), measured trajectory of local THz temporal waveform (the position labelled red in SOP (**e**)).

Similarly, the experimental results of THz vortex beam while the phase delay was set as -π/4 are shown in Fig. 6. The amplitude distributions in Fig. 6(a) and (b) show similar elliptical phase distribution as in Fig. 5, except the phase difference between the pump beams being of opposite sign would correspond to counterclockwise local field vector rotation. The $E_y$ has maximum amplitudes along horizontal and vertical directions, while $E_x$ has minimum amplitudes along them. The evolutions of wavefront of the two eigenstates over 1.65 ps (50 frames) are also attached in videos (Fig. 6_Ex.mp4 and Fig. 6_Ey.mp4). In this case, the phase distributions in Fig. 6(c) and (d) implies the TC of +2. The 3D reconstructed THz electric field trace in Fig. 6(f) demonstrates the left-handed elliptical polarization with long axis along $E_x$ axis.

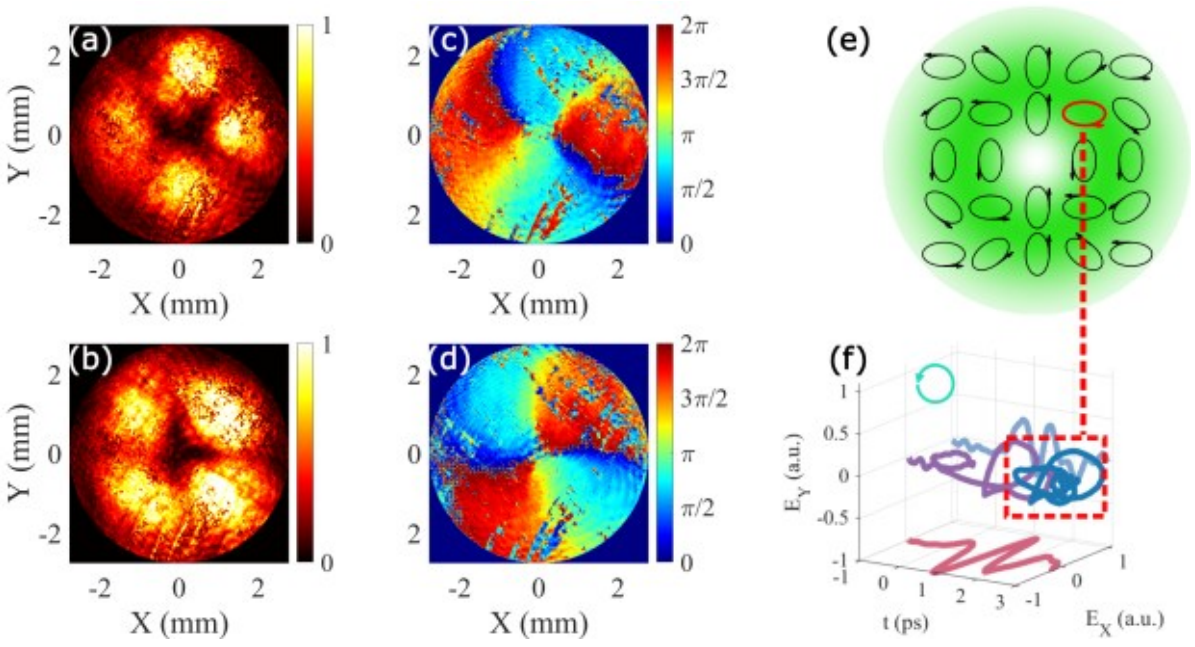

**Fig. 6. Measurement of left-handed elliptical THz vortex beam.** (**a**), THz electric field amplitude distribution along $E_y$ axis. (**b**), Amplitude distribution along $E_x$ axis. (**c**), Phase distribution along $E_y$ axis. (**d**), Phase distribution along $E_x$ axis. Color bars in (**a**) and (**b**) show the intensities in arbitrary units, and color bars in (**c**) and (**d**) represent the phase scale (radian). (**e**), Simulated trajectory of THz temporal waveform. (**f**), measured trajectory of local THz temporal waveform (the position labelled red in SOP (**e**)).

The phase and amplitude of the CV THz beam, representing the state Fig.1-**c** on the equator with zero phase delay between pumps, was also measured and reconstructed, confirming the validity of the transformation along the sampled meridian. The five measured states ($\varphi$ = 0, $\pm\pi/4$, $\pm\pi/2$) span the meridian from pole to pole, and any intermediate $\varphi$ is realized by the same phase-delay control demonstrated in Fig. 4. The experimental results of the CV THz beam are illustrated in Fig. 7. The amplitudes in Fig. 7(a) and (b) exhibit the clear four-petal feature, with the phases between two lobes have a $\pi$ shift. This verifies the CV beam structure. The measured trajectory of THz waveform shows a linear polarization along $E_x$ axis, that matches the SOP simulated in Fig. 7(e). In contrast to the circularly or elliptically polarized vector vortex beams, the transversal wavefronts in this case do not rotate as time progresses (see Fig. 7_Ex.mp4 and Fig. 7_Ey.mp4). This indicates that the net OAM of this and other equatorial states is equal to zero. Those states are sometimes called maximally nonseparable angular momentum states [10].

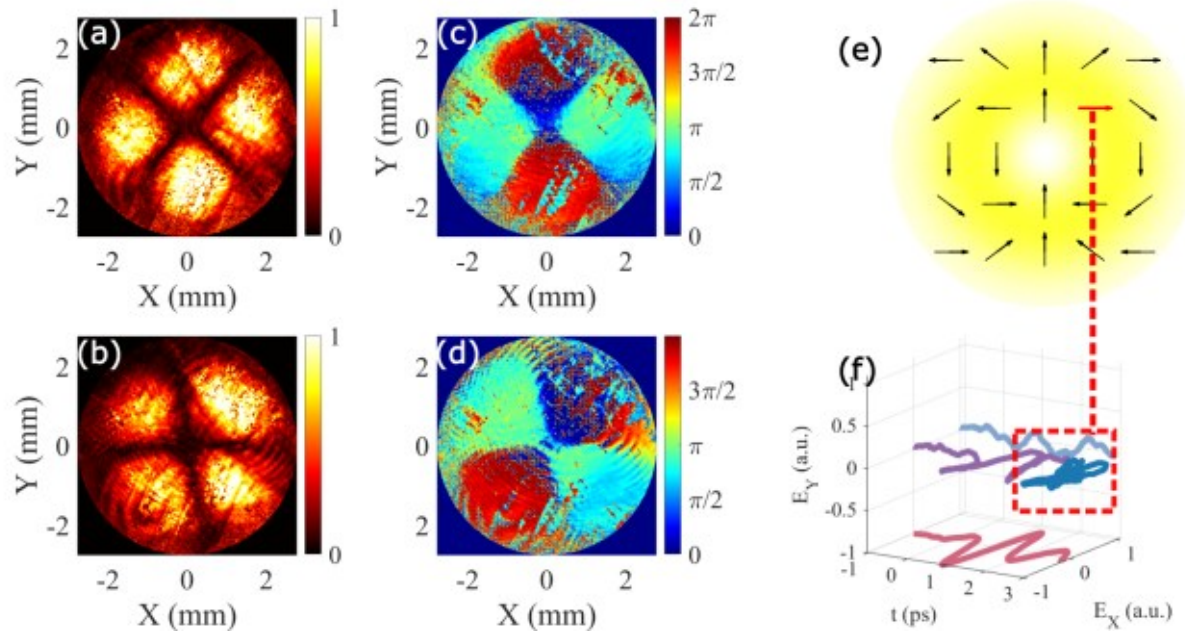

**Fig. 7. Measurement of CV THz beam with S-waveplate at 0°.** (**a**), THz electric field amplitude distribution along $E_y$ axis. (**b**), Amplitude distribution along $E_x$ axis. (**c**), Phase distribution along $E_y$ axis. (**d**), Phase distribution along $E_x$ axis. Color bars in (**a**) and (**b**) show the intensities in arbitrary units, and color bars in (**c**) and (**d**) represent the phase scale (radian). (**e**), Simulated trajectory of THz temporal waveform. (**f**), measured trajectory of local THz temporal waveform (the position labelled red in SOP (**e**)).

To further verify another manipulating degree of freedom, that is the rotation angle of S-waveplate, another state on the equator was generated by rotating the S-waveplate from 0° to 90°, with phase delay remaining as zero. The amplitude and phase profiles of this CV THz beam are shown in Fig. 8(a)-(d) (see wavefront evolution in Fig. 8_Ex.mp4 and Fig. 8_Ey.mp4), exhibiting the same four-petal feature as Fig. 7(a)-(d). The 3D trace was also reconstructed in Fig. 8(f), corresponding to the local polarization labelled in red in Fig. 8(e). This CV beam agrees well with the SOP of Fig.1-**g**, as we expected. The vector vortex THz beams at the equator have been verified to transform via S-waveplate rotation, confirming that THz vortex beams at the same latitude can be correctly generated by rotating S-waveplate, as predicted by our theory. By adjusting the pulse delay between pumps and the rotation angle of the S-waveplate as two degrees of freedom, six representative SOPs on the HOP sphere were generated and characterized, spanning the linear, elliptical, and circular polarization regimes predicted by the two-parameter mapping.

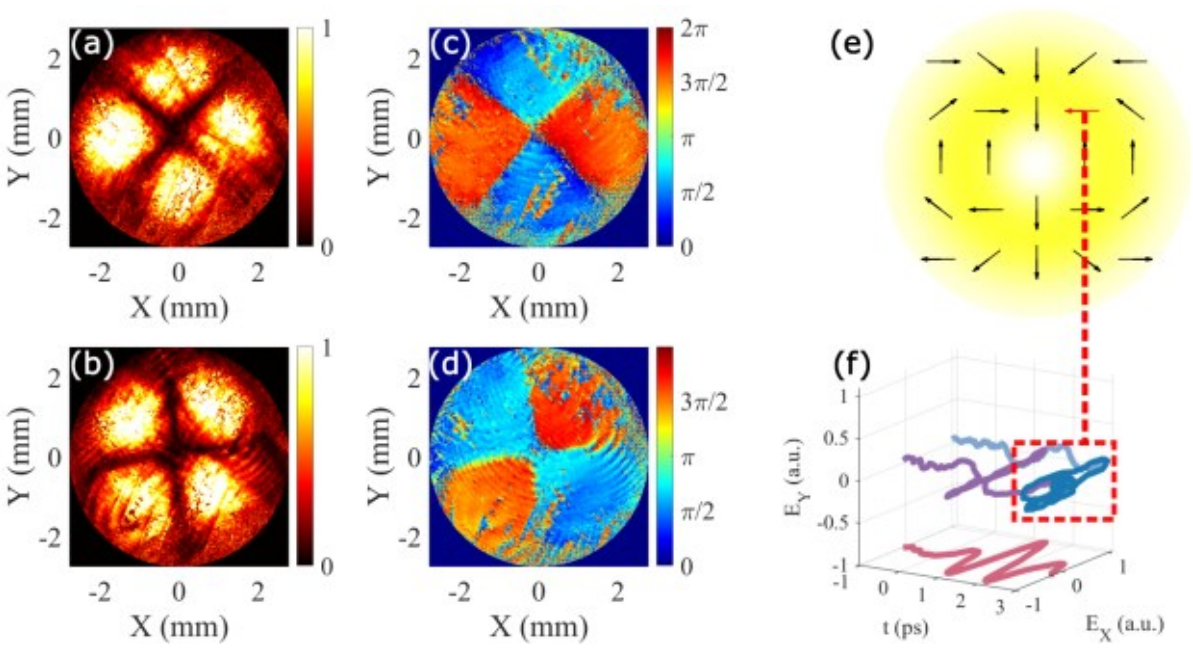

**Fig. 8. Measurement of CV THz beam with S-waveplate at 90°.** (**a**), THz electric field amplitude distribution along $E_y$ axis. (**b**), Amplitude distribution along $E_x$ axis. (**c**), Phase distribution along $E_y$ axis. (**d**), Phase distribution along $E_x$ axis. Color bars in (**a**) and (**b**) show the intensities in arbitrary units, and color bars in (**c**) and (**d**) represent the phase scale (radian). (**e**), Simulated trajectory of THz temporal waveform. (**f**), measured trajectory of local THz temporal waveform (the position labelled red in SOP (**e**)).

**Topological transformation.** To gain further insight into the underlying mechanism, we focus on the equatorial states on the HOP sphere, defined by $\varphi$ = 0. These states form a complete basis of the HOP-sphere mapping. In principle, any state on the sphere is reached by combining one equator basis state with the phase-delay control demonstrated in Figs. 3–7, so the method spans the full sphere by construction. Within this framework, the global phase of the pump vortex pair governs the spiral phase (OAM) of the generated THz field, while the relative phase determines the local polarization state (SAM).

The synthesis of representative equatorial states (**c**, **f**, **g**, and **h** in Fig. 1.) from two linearly polarized near-infrared pump beams is illustrated schematically in Fig. 9. These equatorial states correspond to $\varphi$ = 0. By rotating the S-waveplate orientation $\theta$ = 0°, 45°, 90°, and 135°, distinct CV pump beams are generated, which are subsequently transformed into the corresponding THz states via OR in [111]-ZnTe. Two CV THz beams with orthogonal local polarizations are generated independently by two pumps, respectively. The resulting THz fields arise from the coherent superposition of two THz beams, leading to well-defined vector vortex states on the equator of the HOP sphere.Fig.1.

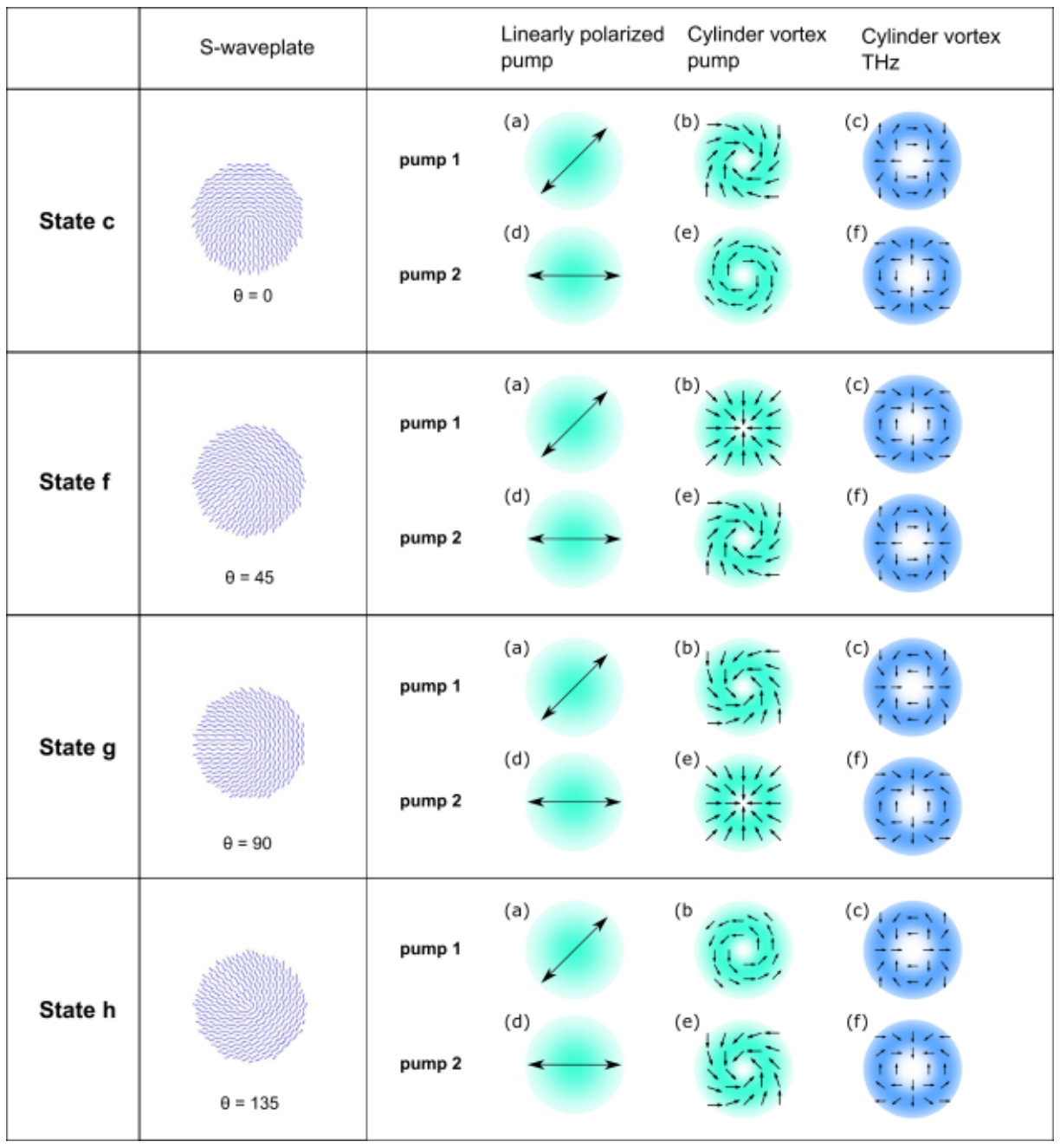

**Fig. 9**. Schematic topological transformation of equatorial states on the HOP sphere. The polarization states of two pump beams and the corresponding THz fields are shown for four representative equatorial states (**c**, **f**, **g**, and **h** in Fig. 1), obtained by setting the S-waveplate orientation to θ = 0°, 45°, 90°, and 135°, respectively. (a, d) Linearly polarized pumps before S-waveplate. (b, e) Corresponding CV pump beams after S-waveplate, carrying polarization winding number of 1. (c, f) Corresponding THz CV beams generated after [111]-ZnTe crystal. All polarization states are presented in the laboratory coordinate frame (X, Y).

We consider the reference case at θ = 0° and φ = 0°, corresponding to state **c** on the equator, which is the baseline state of the mapping framework. The measured pump and THz fields of this case are shown in Fig. 10. The THz fields generated independently by the two CV pumps, THz 1 and THz 2 as shown in Fig. 10**b**, display four-petal profiles, indicating CV structure with a polarization rotation of 4π (corresponding to 2π for the pump of which the intensity profiles shown in Fig. 10**a**). The local polarizations of the two THz fields are mutually orthogonal at all azimuthal positions, representing two orthogonal eigenstates of the vector vortex mode. Their coherent superposition THz produces a resultant CV beam with the same polarization winding number and a rotated local polarization distribution.

The orthogonality of the two THz eigenstates forms a natural basis for the vector vortex mode, enabling the synthesis of arbitrary polarization states via controlled phase delay and thereby providing access to states along the meridians of the HOP sphere. Taken together, the experimental evidence covers the HOP sphere on two independent layers. The THz amplitude and phase profiles are measured along the meridian through equator state c (Figs. 3–7) and at a second equator state (Fig. 8), while the underlying topological transformation, that from pump vortex structure to individual THz vortex structures to the composite THz field, is verified independently for state **c** in Fig. 10. The remaining equator states are connected to the same transformation analysis in Fig. 9, and any non-equatorial state is reached by continuous φ tuning from its corresponding equatorial basis state.

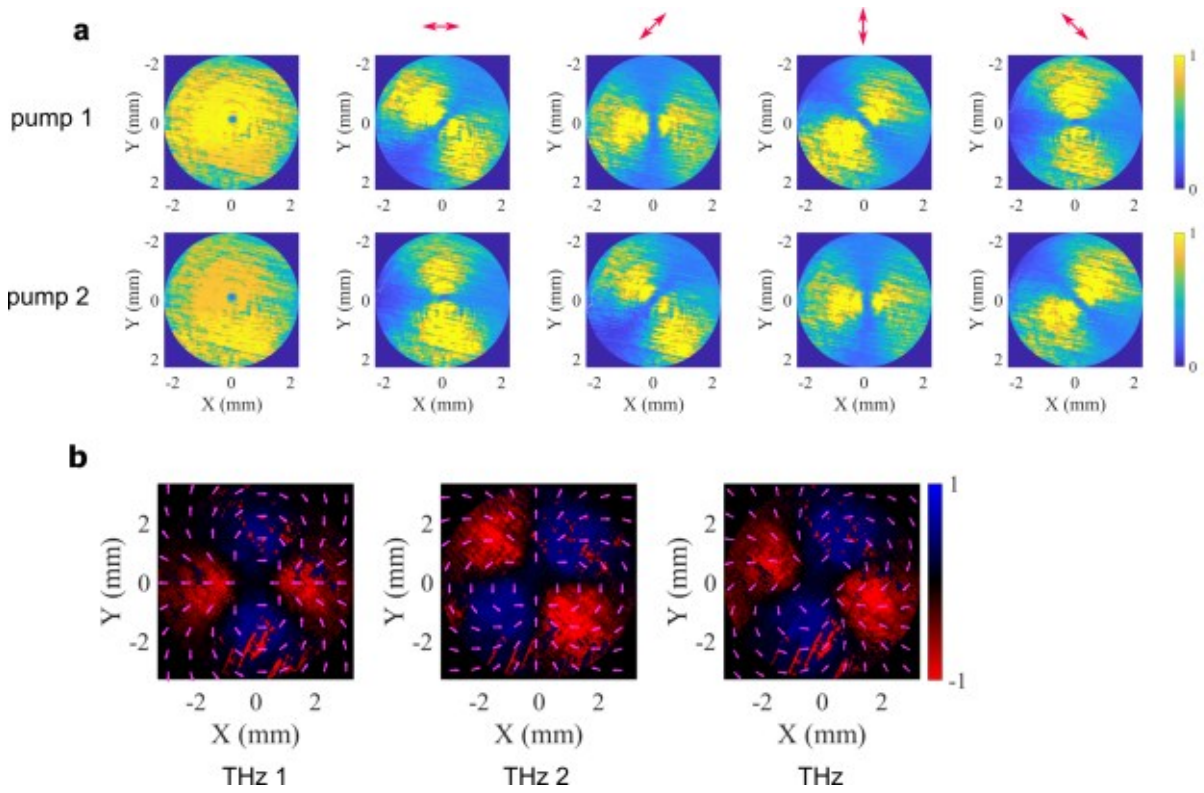

**Fig. 10.** Experimental verification of topological transformation of the baseline case of state **c**. **a** Measured intensity distributions of the pump beams after a 0° S-waveplate, recorded without and with a polarizer. The upper and lower rows correspond to Pump 1 and Pump 2, respectively. Red arrows indicate the transmission orientation of the polarizer, revealing the local polarization structure of the CV beams. **b** Normalized electric field distributions of the corresponding THz fields. THz 1 and THz 2 are generated independently by Pump 1 and Pump 2, respectively, and their coherent superposition forms the resulting THz beam. Purple arrows denote the local polarization orientations. All THz cases exhibit a polarization winding number of 2 around the central singularity. The fields are presented in the laboratory coordinate frame (X, Y).

## Discussion

In conclusion, we demonstrate a versatile approach for synthesizing arbitrary vector-vortex THz beams via OR driven by dual infrared pump beams. By independently controlling the phase delay between the pump pulses and the rotation angle of the S-waveplate, we establish a two-parameter mapping that spans the full HOP sphere by construction. Based on this scheme, we experimentally realize and characterize representative THz HOP states across linear, circular, and elliptical polarization regimes — along a sampled meridian and on the equator — in good agreement with theoretical predictions, verifying the capability of the method to access arbitrary states on the sphere. The results reveal a well-defined topological transformation from the pump configuration to the generated THz fields, arising from the controlled combination of SAM and OAM. Beyond the |l| = 2 sphere demonstrated here, this approach can be extended to higher-order HOP spheres by tailoring the topological winding number of the pump beams. The order of the resulting HOP sphere is twice the winding number of the pump beams, as governed by the OR process. This scheme can also be generalized to other zinc-blende crystals, such as GaP and GaAs, with appropriate phase-matching pump wavelengths. In addition, the use of OR enables straightforward power scaling, suggesting the potential for generating high-field structured THz beams. Overall, this approach provides a flexible and efficient nonlinear platform for tailoring structured THz fields with arbitrary states on the HOP sphere, opening new opportunities for applications in THz imaging, spectroscopy, and light-matter interaction.

## Methods

### THz Generation

**Experimental parameters.** A horizontally linearly polarized laser with a central wavelength of 800 nm, a pulse duration of 33 fs, a repetition rate of 1 kHz (Astrella, Coherent) was used as pump and probe beam. A 111-cut nonlinear ZnTe crystal with a thickness of 1.5 mm is used for THz generation via OR. The generated THz pulses are few-cycle with durations on the order of several ps.

**Control of pump polarization and phase delay.** The pump beam is split into two arms (Pump 1 and Pump 2) with independently controlled polarization and relative phase delay. The polarization of Pump 1 is adjusted using a half-wave plate and a polarizer, while Pump 2 remains horizontally linearly polarized. The two pump beams are 45° linearly polarized relative to each other with balanced power of 30 mW. An S-waveplate is inserted in the common beam path before the nonlinear crystal to convert the linearly polarized pump beams into cylindrical vector beams. Both pump beams pass through the same S-waveplate, ensuring identical spatial mode conversion. The rotation angle θ of the S-waveplate, defined with respect to the horizontal polarization axis, is used to control the azimuthal parameter of the generated THz field. A motorized delay line is applied to Pump 2 to introduce a controllable temporal delay with fs resolution, covering a full 2π phase range at the target THz frequency.

### THz Detection

**Detection scheme.** The generated THz field is detected using electro-optic sampling in a 110-ZnTe crystal with a thickness of 250 μm. The spatial THz electric field distribution is recorded by a synchronized CCD camera using a dynamic subtraction technique [32]. The power of the probe beam used in this paper was 6 mW as dictated by the CCD camera properties. Two 4f imaging systems with a total demagnification of 4 are employed for THz beam imaging.

**THz polarimetry technique.** THz polarimetry technique is built by inserting a pair of wire-grid polarizers (WGP) before the detection crystal, as illustrated in Fig. 11. The orientation of the second WGP (WG2) has a fixed orientation, transmitting horizontal polarization, parallel to the laboratory axis X and to the direction of the [110]-cut ZnTe detection crystal, thus in the direction of maximum electrooptic response. The first WGP (WG1) is set at ±45° with respect to WGP2 to measure the components Ex and Ey of THz electric field, separately. The orientations of Ex and Ey that we call the field reconstruction frame and the laboratory coordinate frame (X, Y) are illustrated in Fig. 11. The THz electric field components are measured without rotating the detection crystal to avoid inconsistency during the measurements.

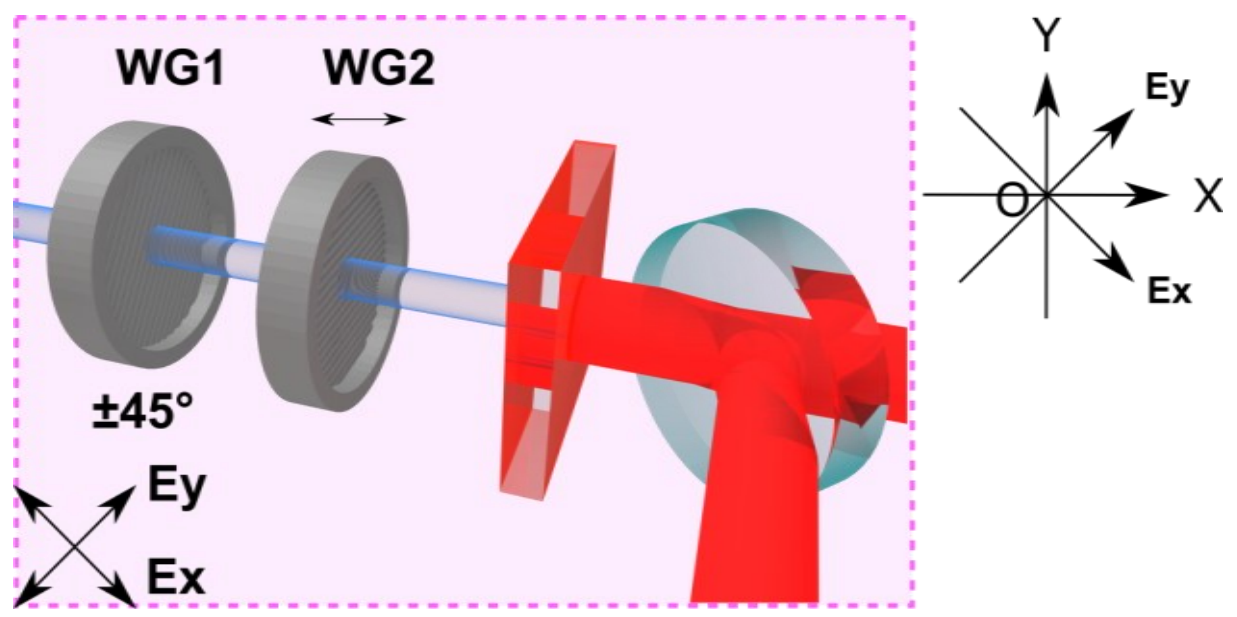

**Fig. 11.** THz polarimetry setup. WG1, WG2: wire grid polarizer, black arrows: transmittance orientation of wire grid polarizers. OXY and OExEy represent laboratory and reconstruction coordinate frames, respectively.

**Zero phase delay calibration.** The zero phase delay between the two pump beams was calibrated using two delay lines (see Fig. 2**b**). First, the Pump 2 arm was blocked, and the temporal THz signal generated by Pump 1 was recorded by scanning the probe-beam delay line. The probe delay was then fixed at the peak position of the THz waveform. Subsequently, both pump beams were unblocked, and the delay line in the Pump 2 arm was adjusted to maximize the THz signal amplitude. This position was defined as the zero-phase delay (φ = 0), corresponding to the temporal overlap of the THz fields generated by the two pump pulses. The S-waveplate was not used during this calibration procedure. After calibration, the S-waveplate was inserted and aligned to the desired orientation angle θ for subsequent measurements.